\documentclass{article}

\usepackage{PRIMEarxiv}

\usepackage[utf8]{inputenc} 
\usepackage[T1]{fontenc}    
\usepackage{hyperref}       
\usepackage{url}            
\usepackage{booktabs}       
\usepackage{amsfonts}       
\usepackage{nicefrac}       
\usepackage{microtype}      
\usepackage{lipsum}
\usepackage{fancyhdr}       
\usepackage{graphicx}%
\usepackage{multirow}%
\usepackage{amsmath,amssymb,amsfonts}%
\usepackage{amsthm}%
\usepackage{mathrsfs}%
\usepackage[title]{appendix}%
\usepackage{xcolor}%
\usepackage{textcomp}%
\usepackage{manyfoot}%
\usepackage{booktabs}%
\usepackage{algorithm}%
\usepackage{algorithmicx}%
\usepackage{algpseudocode}%
\usepackage{listings}%
\newcommand{\vx}{{\mathbf{x}}}
\newcommand{\vy}{{\mathbf{y}}}
\newcommand{\vu}{{\mathbf{u}}}
\newcommand{\vv}{{\mathbf{v}}}
\newcommand{\bbR}{\mathbb{R}}
\newcommand{\diag}{\text{diag}}

\newcommand{\eref}[1]{Eq.~\eqref{eq:#1}}
\newcommand{\fref}[1]{Fig.~\ref{fig:#1}}

\newcommand{\aref}[1]{\algo.~\ref{algo:#1}}

\newcommand{\algo}{\textbf{Algorithm}}

\def\figclip#1#2#3{\includegraphics[trim=#1, clip, width=#2\columnwidth]{#3}}
\title{Uncertainty-aware t-distributed Stochastic Neighbor Embedding for Single-cell RNA-seq Data
}

\author{
  Hui Ma, \\
  School of Mathematics and Statistics \\
  Central South University \\
  \texttt{huima333@gmail.com} \\
   \And
  Kai Chen \\
  School of Mathematics and Statistics \\
  Central South University \\
  \texttt{kaichen6@csu.edu.cn} \\
}

\begin{document}
\maketitle

\begin{abstract}
Nonlinear data visualization using t-distributed stochastic neighbor embedding (t-SNE) enables the representation of complex single-cell transcriptomic landscapes in two or three dimensions to depict biological populations accurately. 
However, t-SNE often fails to account for uncertainties in the original dataset, leading to misleading visualizations where cell subsets with noise appear indistinguishable. To address these challenges, we introduce uncertainty-aware t-SNE (Ut-SNE), a noise-defending visualization tool tailored for uncertain single-cell RNA-seq data. By creating a probabilistic representation for each sample, Our Ut-SNE accurately incorporates noise about transcriptomic variability into the visual interpretation of single-cell RNA sequencing data, revealing significant uncertainties in transcriptomic variability. Through various examples, 
we showcase the practical value of Ut-SNE and underscore the significance of incorporating uncertainty awareness into data visualization practices. 
This versatile uncertainty-aware visualization tool can be easily adapted to other scientific domains beyond single-cell RNA sequencing, making them valuable resources for high-dimensional data analysis.
\end{abstract}

\section{Introduction}\label{sec1}
Visual exploration of high-dimensional data is crucial for the comprehensive analysis of single-cell datasets. Fluorescence, mass, and sequencing-based cytometry require tools capable of identifying the combinations of proteomic and/or transcriptomic markers that define complex and diverse cell phenotypes within mixed populations. 
It has proven particularly valuable in the rapidly growing field of single-cell transcriptomics. High-throughput single-cell RNA sequencing (scRNA-seq) technologies allow researchers to study gene expression with unprecedented resolution across a wide range of tissues, organisms, and biological conditions \cite{svensson2018exponential,ranek2023feature}.
Driven by the high dimensionality and large scale of scRNA-seq data, unsupervised visualization plays a pivotal role in comprehensively studying cell heterogeneity across diverse biological tissues and conditions, providing them with a profound glimpse into the intricacies of cell biology. Most researchers rely on visualization techniques that project scRNA-seq datasets into two- or three-dimensional space, revealing underlying patterns such as clusters or trajectories, and ultimately advancing our understanding of complex biological systems and disease mechanisms. 

Several visualization techniques have been applied to scRNA-seq data, with varying degrees of success. Linear methods, such as principal component analysis (PCA) \cite{mackiewicz1993principal,kurita2019principal}, are often unsuitable for visualizing scRNA-seq data because they fail to capture the nonlinear relationships present. In contrast, t-Distributed Stochastic Neighbor Embedding (t-SNE) \cite{cai2022theoretical,van2008visualizing} is a cutting-edge algorithm designed for nonlinear data representation. It generates a low-dimensional distribution, or "map," of high-dimensional data, where distinct clusters of points, or "islands," represent observations that are similar in the original high-dimensional space. 
This property is beneficial in clustering cells with similar gene expression profiles, reflecting their biological similarity in terms of overall structure and heterogeneity. 
Additionally, t-SNE maps are used to categorize single-cell data into biologically relevant populations for downstream quantification, achieved through expert-guided gating or unsupervised clustering. Visualizations generated by other nonlinear embedding algorithms, such as locally linear embedding (LLE) \cite{roweis2000nonlinear, li2008locally}, multidimensional scaling (MDS) \cite{carroll1998multidimensional,saeed2018survey,hagele2022uncertainty}, self-organizing map (SOM) \cite{wickramasinghe2019deep}, and uniform manifold approximation and projection (UMAP) \cite{mcinnes2018umap,becht2019dimensionality}, Laplacian eigenmaps \cite{belkin2003laplacian, levin2016laplacian}, Harmony, and Scanorama can be interpreted in much the same way. These methods have been instrumental in uncovering hidden patterns and structures within single-cell genomics data, providing valuable insights for researchers in various fields such as biology, medicine, and biotechnology. However, handling uncertain data presents significant challenges that require additional consideration. Unfortunately, visualization techniques like t-SNE and UMAP do not inherently provide a mechanism for addressing uncertainty or propagating it through the reduction process.

In this study, we focus on t-SNE due to its widespread use as a nonlinear visualization method. However, despite its strengths, t-SNE has a significant limitation: it overlooks the uncertainty inherent in the original dataset. In other words, the scRNA-seq data contain noise stemming from experimental procedures, technical limitations, and biological variability. This noise contributes to variability in the data, potentially creating spurious clusters or masking the true underlying structure. The standard t-SNE assumes a somewhat naive approach to the data, ignoring the presence of such variability. As a result, critical uncertainty information is lost in the visualizations.
Capturing uncertainty between clusters in the low-dimensional space holds the potential to reveal heterogeneity both between and within clusters (i.e., subclusters). In scRNA-seq studies, this heterogeneity reflects variability in gene expression within cell populations. Therefore, incorporating uncertainty into the visualization could offer a new perspective on biological processes, enriching the interpretation of cluster distribution and providing valuable insights into the transcriptomic landscape of individual cells. 

To address the challenge of visualizing uncertainty, we propose uncertainty-aware t-SNE (Ut-SNE), a novel adaptation of t-SNE for handling uncertain data, designed to help researchers more accurately visualize and gain deeper biological insights from single-cell transcriptomic experiments. Ut-SNE provides a more robust and interpretable visualization of high-dimensional data with noise, offering valuable insights into the underlying structure and variability of the data. 
The key contributions of our paper are as follows: Initially, we extend t-SNE to accommodate uncertain data by developing a versatile and rigorous mathematical framework that is compatible with a wide range of distributions. Next, we refine this general concept for the specific scenario of normally distributed input data by treating the observed value of each data point as mean and augmenting this point with an auxiliary variance term that accounts for the variability of the point. 

We develop a general, integrable measure of uncertain similarity, which intuitively represents the expectation of distance between data points. 
Furthermore, we transform the objective function of t-SNE into a probability space by using the uncertain similarity, which results in an efficient numerical algorithm and a clearer understanding of the mathematical components involved in the t-SNE algorithm.  
Furthermore, we introduce several visualization techniques tailored to our method that address the reliability and sensitivity of dimensionality reduction in the presence of uncertainty. The algorithmic techniques introduced here could be applied to other high-dimensional datasets, making them broadly applicable.

In this work, we evaluate Ut-SNE's performance using standard settings aligned with best practices. 
To showcase the benefits of maintaining uncertainty in data visualizations, we utilize Ut-SNE on an assortment of published scRNA-seq datasets, including samples from human breast tissue, human adipose tissue, immune cells in peripheral blood, and immune cells in tumors. The uncertainty-aware visualization technique not only provides more information than conventional methods but also reveals biological insights that might otherwise be overlooked. Ut-SNE generates better visualization in preserving data structure than standard t-SNE, projects the uncertain information that others miss, and uncovers hidden substructures and the distribution of each cluster.  
Our findings indicate that uncertainty-aware visualization techniques can uncover previously unseen patterns within single-cell transcriptomic data, thereby enhancing our comprehension of biological processes.

\section{Methods}\label{sec13}
\subsection{Framework of t-SNE}
t-SNE \cite{van2008visualizing}, one of the most advanced visualization algorithms today, evolved from its predecessor, a method known as SNE. The core concept of SNE involves characterizing the relationships between pairs of high-dimensional points through normalized affinities. Close neighbors exhibited high affinity, while distant samples had near-zero affinity. SNE then arranged the points in two dimensions, aiming to minimize the Kullback-Leibler (KL) divergence between the affinities in high and low dimensions.  

Specifically, given a high-dimensional dataset $X=[{{\vx}_{1}},{{\vx}_{2}},...,{{\vx}_{n}}]^{\top}$ where $\vx\in{R}^{D}$ and $n$ is the number of samples, t-SNE aims to find the corresponding low-dimensional embedding $Y=[{{\vy}_{1}},{{\vy}_{2}},...,{{\vy}_{n}}]$ for $X$, where $\vy\in {{R}^{d}}$ and  $d<D$. Within t-SNE, if samples $\vx_{i}$ and $\vx_{j}$ are close in the high-dimensional space, their low-dimensional embeddings $\vy_{i}$ and $\vy_{j}$ should be close. 
The likeness between samples $\vx_{i}$ and $\vx_{j}$ is expressed through conditional probability $p_{j|i}$. This probability signifies the probability that $\vx_{i}$ would choose $\vx_{j}$ as its neighbor if neighbors were selected based on their probability density under a Gaussian centered at $\vx_{i}$. In the case of closely situated samples,  $p_{j|i}$ tends to be relatively high, while for significantly distant samples, $p_{j|i}$ becomes nearly infinitesimal.

In t-SNE, we define the pairwise similarity (denoted by $p_{ij}$ ) between samples $\vx_{i}$ and $\vx_{j}$ in the high-dimensional space as
\begin{align} \label{eq:tpij} 
p_{ij} &= \frac{p_{j|i} + p_{i|j}}{2n},
\end{align}
where
\begin{align} \label{eq:pji}
p_{j|i} &= \frac{\exp \big(-\frac{\|\vx_{i} - \vx_{j}\|^2}{2\sigma_i^2}\big)}{\sum\limits_{k\ne i} \exp \big(-\frac{\|\vx_{i} - \vx_k\|^2}{2\sigma_i^2}\big)}.
\end{align}

A crucial parameter involved in \eref{pji} is the variance $\sigma_{i}$ of the Gaussian centered at each high-dimensional data point $\vx_{i}$. Since data density varies across the dataset, it is important to choose a reasonable value for $\sigma_{i}$ \cite{kobak2019art}. 
Typically, we select the variance $\sigma_{i}$ by presetting the perplexity of the distribution as follows:
\begin{align}
	\mathrm{Perp}({P}_{i})={2}^{H({P}_{i})},
\end{align}
where $H({{P}_{i}})=-\sum_{j}{p}_{j|i}\log_{2}{p}_{j|i}$ is the Shannon entropy of $P_{i}$ measured in bits.

The low-dimensional counterparts, $\vy_i$ and $\vy_j$, corresponding to the high-dimensional samples $\vx_i$ and $\vx_j$, allow the calculation of a similar conditional probability $q_{j|i}$. Thus, we characterize the similarity between $\vy_i$ and $\vy_j$ by using the t-distribution with one degree of freedom:
\begin{align}\label{eq:qji}
	q_{ij} &= \frac{\big(1+{\|\vy_{i} - \vy_{j}\|^2}\big)^{-1}}{\sum\limits_{k\ne l}\big(1+{\|\vy_{k} - \vy_l\|^2}\big)^{-1}}.
\end{align}
If the low-dimensional points $\vy_i$ and $\vy_j$ accurately represent the similarity between the high-dimensional samples $\vx_i$ and $\vx_j$, the conditional probabilities $p_{ij}$ and $q_{ij}$ will be identical. Inspired by this insight, t-SNE discovers a low-dimensional data embedding that minimizes the discrepancy between $p_{ij}$ and $q_{ij}$. A natural measure of how faithfully $q_{ij}$ models $p_{ij}$ is the KL divergence across all samples. The cost of t-SNE for minimizing the KL divergences is defined as follows:
\begin{align}\label{eq:t cost}
\begin{split}
    \mathcal{L}=\sum\limits_{i}\mathrm{KL}({P}_{i}\|{Q}_{i})=\sum\limits_{i, j}{p}_{ij}\log \frac{p_{ij}}{q_{ij}}.
\end{split}
\end{align}
We further obtain the minimization by using a gradient descent approach (in \aref{tsne}). This minimization significantly punishes employing widely separated $\{\vy_{i}, \vy_{j}\}$ to represent nearby samples $\{\vx_{i}, \vx_{j}\}$.

\begin{algorithm}[!h]
\renewcommand{\algorithmicrequire}{\textbf{Input:}}
\renewcommand{\algorithmicensure}{\textbf{Output:}}
\caption{Standard t-SNE} \label{algo:tsne}
\begin{algorithmic}[1]
\Require  high dimensional data $X$, 
number of iterations $T$, learning rate $\eta$, and momentum $\alpha(t)$. 
    \Ensure low dimensional embedding $Y^{(T)}$.
    \State Compute pairwise affinities $p_{j|i}$ with $\mathrm{Perp}$;
    \State Set $p_{ij}=\frac{p_{i|j}+p_{j|i}}{2n}$;
    \State Initialize $Y^{(0)}=[\vy_{1}, \vy_{2}, \dots ,\vy_{n}]^{\top}$ using PCA;
    \For{$t=1$ to $T$}
    \State Compute low-dimensional affinities $q_{ij}$;
    \State Compute gradient $\frac{\partial \mathcal{L}}{\partial \vy}$;
    \State Set ${{\vy}^{(t)}}={{\vy}^{(t-1)}}+\eta \frac{\partial \mathcal{L}}{\partial \vy}+\alpha (t)({{\vy}^{(t-1)}}-{{\vy}^{(t-2)}})$.
    \EndFor
    \end{algorithmic}
\end{algorithm}

\subsection{Uncertainty-aware t-SNE}
To incorporate and deal with uncertain high-dimensional data, we introduce Ut-SNE which can generate a low-dimensional embedding with uncertainty. Particularly, the embedding not only captures the underlying structure of the data but also accounts for the associated uncertainty. In this section, we delve into the rationale and derivation of a universal Ut-SNE capable of handling a wide array of distribution types. We zoom in on a particular form of Ut-SNE that's both typical and significant: uncertainty characterized by Gaussian distributions. 

\textbf{General probabilistic representation of high-dimensional distance}: 
For uncertainty-aware t-SNE, the measure of distance plays an important role because it reflects the uncertainty of the samples. 
On the other hand, both Gaussian and Cauchy kernels in t-SNE imply that the points located from $\vx_{i}$ contribute less in the $p_{\centerdot |i}$. Based on the definition of $p_{\centerdot |i}$, the distance $d_{\centerdot i}$ at point $\vx_{i}$ determines whether the points are omitted from the conditional distribution $p_{\centerdot i}$.  

In order to efficiently measure the pairwise distance between uncertain data points,  we propose a probabilistic representation for the distance with data variability. Given two data points $\vx_{i}, \vx_{j}$,  we parameterize their distance as a general distance function $d(\vx_{i}, \vx_{j})$. We then define a joint distribution $f(\vx_{i}, \vx_{j})$ that reflects the probability of the distance due to the variability of data points. If the the locations of $\vx_{i}$ and $\vx_{j}$ are appeared with low probability, $f(\vx_{i}, \vx_{j})$ is lower. 
Therefore, we have the general probabilistic representation of distance incorporating data uncertainty as:
\begin{align}\label{eq:high-expectation-o}
\mathbb{E}[\| \vx_{i} - \vx_{j} \|^{2}] &= \iint_{\bbR^{D}\times \bbR^{D}} f(\vx_{i}, \vx_{j}) d(\vx_{i}, \vx_{j}) d\vx_{i}\,d\vx_{j}. 
\end{align}

The integration is performed across the space of these pairs.
Due to the $i.i.d$ of data,  $\vx_{i}$ is independent with $\vx_{j}$ and we have $f(\vx_{i}, \vx_{j}) = f(\vx_{i})f(\vx_{j})$. 

\textbf{Gaussian distribution of data point}: While the general probabilistic representation of distance above provides the framework for data points following arbitrary distributions, we aim to derive a particular and typical form based on Gaussian distributions. This choice is due to the tractability of the integrals, which allows us to handle the integrals analytically. We consider $\vx_{i} \sim  \mathcal{N}(\vx_{i};\vx^{*}_{i}, {S}_{i})$, where $\vx^{*}_{i}\in \bbR^{D}$ and $S_{i} \in \bbR^{D\times D}$ denotes the mean and covariance of data points $\vx_{i}$, respectively. 
Once we specify the Gaussian distribution for one data point, another key factor is choosing the type of covariance used to model the spread of each data point. The covariance type can significantly affect the performance of the Ut-SNE in terms of its ability to accurately represent the underlying distribution of the data. There are some common types of covariance for high dimensional Gaussian distribution, including spherical, diagonal, and full covariance. 

Spherical covariance assumes that the covariance matrix is proportional to the identity matrix, meaning each feature of a data point has the same variance and there is no correlation between features. This approach is computationally efficient but can be too restrictive for data with complex structures, such as when different dimensions have varying variances. Diagonal covariance, on the other hand, models the variance along each axis while still assuming no correlation between features, meaning the off-diagonal elements are zero. This provides a balance between computational efficiency and flexibility compared to spherical covariance.
A full covariance matrix allows each feature to form its own arbitrary ellipsoid (or hyperellipsoid in higher dimensions), capturing both variances and correlations between features. While this offers the greatest flexibility in modeling the data, it comes at the cost of higher computational complexity and a greater risk of overfitting, especially in high-dimensional spaces, as more parameters need to be estimated

Here, we start with diagonal covariance due to its balance between flexibility and computational efficiency. Therefore, the matrix $\diag(S_{i})$ implies that all dimensions of data points are independent.
Furthermore, the general probabilistic representation of distance 
can be written as:
\begin{align}\label{eq:high-expectation}
\begin{split}
\mathbb{E}[\|\vx_{i}-\vx_{j}\|^{2}] &= \iint_{\bbR^{D}\times\bbR^{D}} \mathcal{N}(\vx_{i};{\vx^{*}_{i}},{{S}_{i}}) \cdot \mathcal{N}(\vx_{j};{\vx^{*}_{j}},{{S}_{j}})\cdot\|\vx_{i}-\vx_{j}\|^{2}\,d\vx_{i}\,d\vx_{j} \\
&= \iint_{\bbR^{D}\times \bbR^{D}} \frac{\exp{\big(\frac{-\vu^{\top}\vu-\vv^{\top} \vv}{2}\big)}}{\sqrt{(2\pi)^D }}\left\|S_{i}^{\frac{1}{2}}\vu - S_{i}^{\frac{1}{2}}\vv + \vx_{i}^{*} - \vx_{j}^{*}\right\|^{2}d\vu d\vv \\
&= \sum\limits_{i=1}^{D}{{{S}_{i}}} + \sum\limits_{i=1}^{D}{{{S}_{j}}} + \|\vx_{i}^{*}-\vx_{j}^{*}\|^{2},
\end{split}
\end{align}
where $\mathcal{N}(\vx_{i};{\vx^{*}_{i}},{{S}_{i}})=\frac{1}{\sqrt{(2\pi)^D |S_{i}|}}\exp{\big(-\frac{1}{2}(\vx_{i}-\vx_{i}^{*}) {S_{i}}(\vx_{i}-\vx_{i}^{*})^{\top}\big)}$. 
Note that the calculation of $\mathbb{E}[\|\vx_{i}-\vx_{j}\|^{2}]$ can be performed efficiently. 
This result extends the earlier Euclidean distance in standard t-SNE. Euclidean distance can be seen as a normal distribution with a covariance matrix set to zero. Note that for full covariance, the computation of the probabilistic distance involves matrix multiplication and eigenvalue decomposition, which has a complexity of $O(D^3)$. If there are $n$ data points within a dataset, the computation complexity will be $O(n^2D^3)$. 
When we employ full covariance, the numbers of experimental data points and data features mostly range from tens to hundreds due to the high computation complexity. However, when employing diagonal covariance in probabilistic distance, Ut-SNE can perform visualization on dataset with thousands or even tens of thousands of high-dimensional data points.

\textbf{Probabilistic representation of low-dimensional distance}: 
Similarly, we can construct the corresponding distribution for the low-dimensional embedding. 
To realize this in a generic approach, we use local projections $\Phi_{1}, \dots, \Phi_{n}$ to express general relationships between high-dimensional data points $\vx_{1}, \dots, \vx_{n}$ and low-dimensional data points $\vy_{1}, \dots, \vy_{n}$, i.e., $\vy_{i} = \Phi_{i}(\vx_{i})$ for $i = 1, \dots, n$. 
In this context, the overall dimensionality reduction becomes a combination of individual local projections, one for each uncertain input distribution. It’s important to note that the forms of these local projections are not yet determined. Therefore, we have the probabilistic distance in low-dimensional space as
\begin{align}\label{eq:low-expection}
\begin{split}
    \mathbb{E}[\|\vy_{i}-\vy_{j}\|^{2}]= &\mathbb{E}[\|\Phi_{i}(\vx_{i})-\Phi_{j}(\vx_{j})\|^{2}]
\end{split}
\end{align}

Put simply, we approximate the local projections using a finite Taylor series tailored to the distribution \cite{hagele2022uncertainty}. Specifically, we consider an affine transformation \(\Phi: \bbR^D \rightarrow \bbR^d\) of first-order Taylor expansion, given by \(\Phi_{i}(\vx_{i}) = A_i \vx_{i}\), is sufficient for multivariate Gaussian distributions, where variables \(A_i \in \bbR^{d\times D}\) . 
Specifically, for a random vector \(\vx_{i} \sim \mathcal{N}(\vx_{i};{\vx^{*}_{i}},{{S}_{i}})\), the transformed distribution is \(\vy = A \vx_{i} \sim \mathcal{N}(\vy_{i};A\vx^{*}_{i},A{S}_{i}A^{\top})\). So we have the probabilistic distance as 
\begin{align}\label{eq:low-expection}
\begin{split}
    \mathbb{E}[\|\vy_{i}-\vy_{j}\|^{2}]=&\iint_{\bbR^{D}\times \bbR^{D}} \mathcal{N}(\vx_{i};{\vx^{*}_{i}},{{S}_{i}}) \cdot \mathcal{N}(\vx_{j};{\vx^{*}_{j}},{{S}_{j}})\|A \vx_{i}-A \vx_{j}\|^{2} d\vx_{i}d\vx_{j}  \\
    &= \sum{A_i \cdot \diag({S}_{i}) \cdot A_i^T} + \sum{A_j \cdot \diag({S}_{j}) \cdot A_j^T} + \|A_i\vx_{i}^{*}-A_j\vx_{j}^{*}\|^{2}.
\end{split}
\end{align}
These affine transformations preserve the normality of high-dimensional data points, resulting in low-dimensional random vectors that remain normally distributed. Due to these adjustments, the transformation has a clear probabilistic interpretation in terms of the distributions. Therefore, the distance in \eref{low-expection} depends only on the local projections.

\textbf{Divergence between original data and low dimensional embedding}:
The similarity $p_{ij}$ and $q_{ij}$ can now be equivalently expressed as a function of the local projections \(\{A_i\}_{i=1}^{n}\). For Ut-SNE, we can substitute both the high dimensional and low dimensional distance terms in standard t-SNE with probabilistic distances. The new similarities $p_{ij} = \frac{p_{j|i} + p_{i|j}}{2n}$ and $q_{ij}$ can be rewritten as 
\begin{align} \label{eq:pji}
p_{j|i} &= \frac{\exp \big(-\frac{\mathbb{E}[\|\vx_{i}-\vx_{j}\|^{2}]}{2\sigma_i^2}\big)}{\sum\limits_{k\ne i} \exp \big(-\frac{\mathbb{E}[\|\vx_{i}-\vx_{j}\|^{2}]}{2\sigma_i^2}\big)}\\
q_{ij} &= \frac{\big(1+{\mathbb{E}[\|\vy_{i}-\vy_{j}\|^{2}]}\big)^{-1}}{\sum\limits_{k\ne l}\big(1+{\mathbb{E}[\|\vy_{i}-\vy_{j}\|^{2}]}\big)^{-1}}.
\end{align}
The learning process of Ut-SNE aims at minimizing the KL divergences between the new similarities $p_{ij}$ and $q_{ij}$ by using a gradient descent algorithm.  Note that we can perform Ut-SNE efficiently 
on scRNA-seq dataset.

\section{Results}\label{sec2}
\subsection{Overview of Ut-SNE}\label{sec3}
Our Ut-SNE projects uncertain single-cell samples into a target space of lower dimensionality. Ut-SNE achieves the visualization with uncertainty by generating embeddings that maintain both the local structure and uncertainty from the high-dimensional scRNA-seq dataset.  
Within the high-dimensional space, we believe that the uncertainty can be expressed as the distribution of a data point. 
We provide a probabilistic representation of the data point to compute a local distribution for every point. 
In addition, the mean and variance of the distribution are sufficient to capture the uncertainty of a data point \cite{van2021compression}. 
This distribution can measure the probabilistic distance from a point to its closest neighbors, effectively capturing the uncertainty. Ut-SNE focuses on preserving probabilistic distances between neighboring points. 
The pair probabilistic distance is indicated by the expectation of the distance between uncertain samples.
The core objective function of Ut-SNE quantifies how well a distribution of embeddings aligns with the distribution of high-dimensional data points and adjusts the embedding and parameter of uncertainty to maximize this alignment.
Our innovation lies in adding an uncertainty term to the objective function, which measures the probabilistic agreement between the local distribution in the original dataset and the embedding. This ensures that the local structure remains intact within the embedding while also conveying information about local uncertainty. 
We present the Ut-SNE framework in \fref{framework}. The input of the Ut-SNE framework contains a single-cell dataset matrix and a variance matrix encoding the uncertainty of the original dataset. Ut-SNE includes a range of pivotal steps,  such as probabilistic representation of data point and embedding, measurement of uncertain similarity between data points and embedding, the construction of joint distribution in high-dimensional and low-dimensional space to incorporate uncertainty, optimization of low-dimensional embedding as well as parameters of uncertainty,  visualization of the uncertainty.

\begin{figure}[h!]
\centering
\begin{tabular}{p{0.1mm}*{1}{c}}
    & \figclip{0 80 0 60}{0.90}{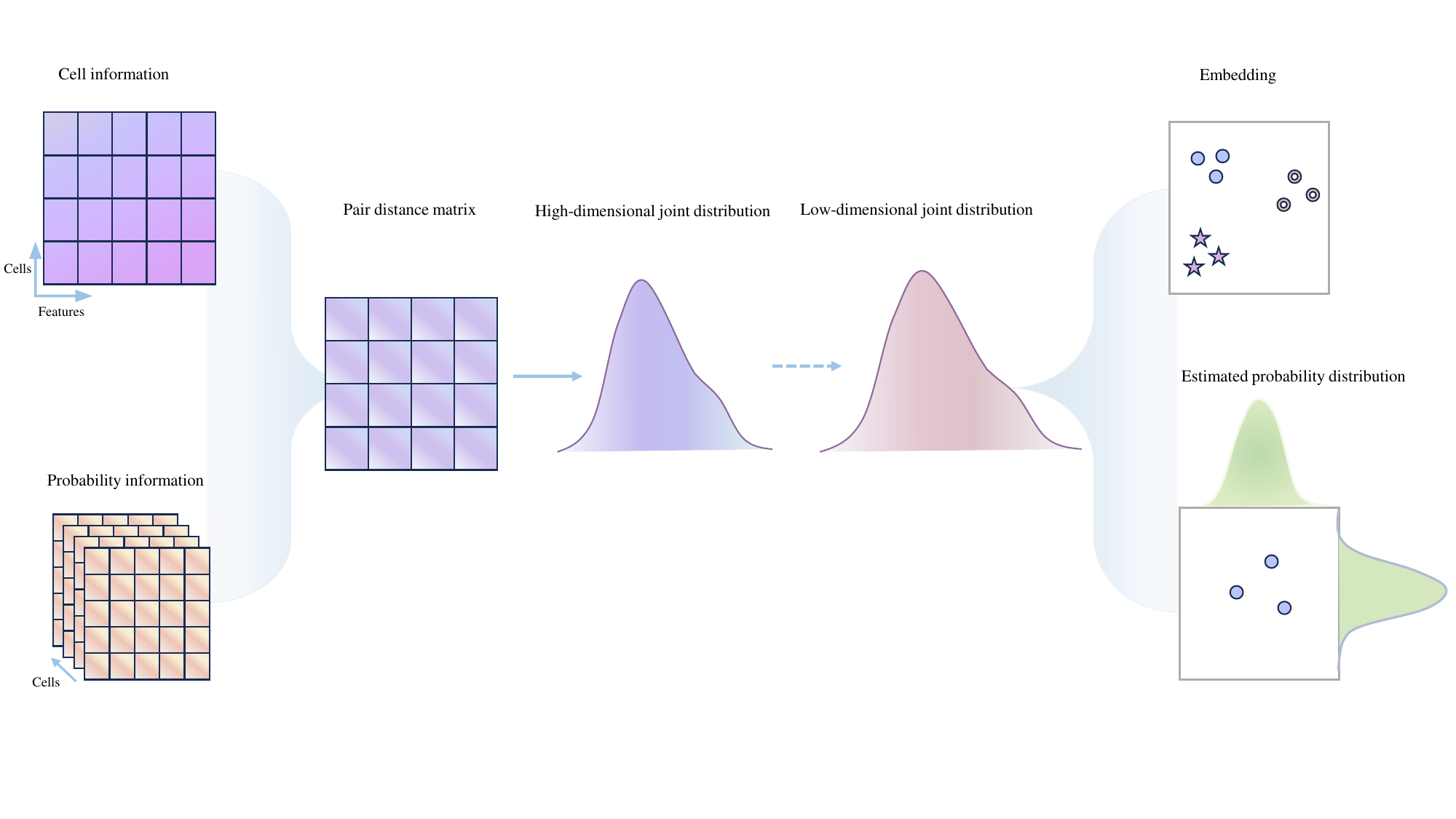}
\end{tabular}
\caption{Overview of Ut-SNE for uncertainty aware visualization: Ut-SNE aims to represent uncertain high-dimensional scRNA-seq data, in 2D or 3D low-dimensional embedding space while maintaining their original uncertainty and structure. Standard t-SNE achieves this by constructing $k$-nearest neighbor graphs to summarize the data manifold. However, t-SNE computes deterministic distances within neighborhoods, which can overlook uncertainty in the original space - a key feature of the data. To address this, we introduce a general, uncertain, integrable measure of similarity. 
The process of Ut-SNE begins with a matrix of cell dataset (denoted by a purple matrix) and a pair probabilistic matrix to calculate the uncertain pair distance matrix, which represents high-dimensional joint distribution. This is then aligned with a low-dimensional joint distribution obtained by optimizing point coordinates to preserve local uncertain distances between neighbors.
}
\label{fig:framework}
\end{figure}

\subsection{Visualization on synthetic datasets with uncertainty}
To highlight the inherent limitations of the standard t-SNE in visualizing the uncertainty of original data, we apply Ut-SNE to synthetic datasets with uncertainty. However, standard t-SNE provides overconfident visual conclusions in this case. Visualizing each Gaussian distributed point with uncertainty, both t-SNE and Ut-SNE project the points in the low-dimensional space, whereas Ut-SNE accurately captures the uncertainty and provides an effective visualization of the uncertainty. We show the visualization results of synthetic datasets in \fref{toy}.
On the other hand, the application of our Ut-SNE to real-world datasets with uncertainty revealed that without incorporating uncertainty, misleading visual conclusions could be drawn. For instance, when visualizing scRNA-seq data points with noise, standard t-SNE produces sharp clusters, whereas Ut-SNE accurately makes a probabilistic distinction between these clusters. When the distribution ranges of the data points overlap, reflecting uncertainty, the lack of incorporation of uncertainty in t-SNE mixes their cluster. 
Our subsequent findings underscore the importance of uncertain modeling in biological analyses.

\begin{figure}[htb]
    \centering
    \scriptsize
    \renewcommand{\tabcolsep}{1.0mm}    
    \begin{tabular}{p{0.5mm}*{2}{c}}
        & \figclip{10 0 10 2}{0.32}{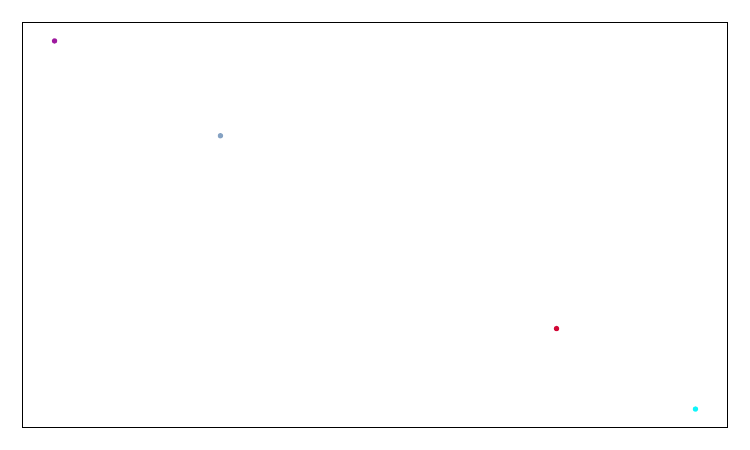} 
        & \figclip{10 0 10 2}{0.32}{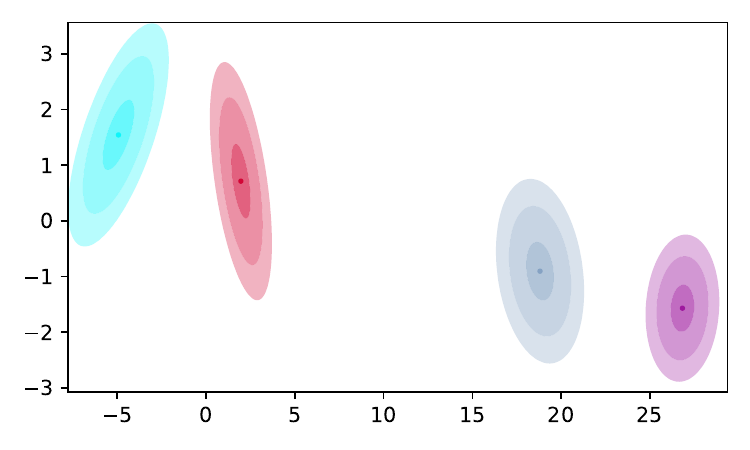}\\
        & (a) & (b) \\
        & \figclip{10 0 10 3}{0.25}{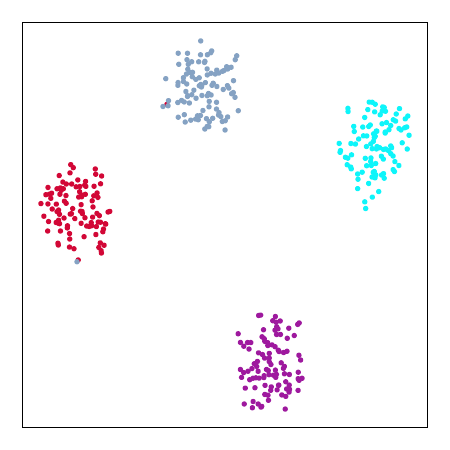} 
         & \figclip{10 0 10 2}{0.38}{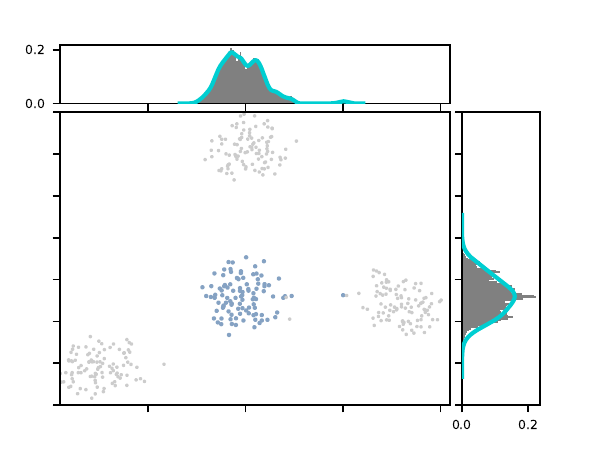}\\
         & (c) & (d) 
    \end{tabular}
    \caption{Uncertainty-aware visualization provides a more accurate representation of the inherent uncertain structure within synthetic datasets when compared to conventional methods. The point clouds of the synthetic dataset are sampled from a mixture of Gaussian in 20 dimensions.  The visualization results of the synthetic dataset include: (a) visualization of four points using standard t-SNE; (b) visualization of four points with different uncertainties (indicated by isolines) using Ut-SNE; (c) visualization of the point clouds using standard t-SNE; (d) 
    visualization of the point clouds and its probability $\frac{1}{n}\sum_{i=1}^{n}\mathcal{N}(\vy_{i}; \vy_{i}^{*}, \mathrm{Var}(\vy_{i}))$ using Ut-SNE. While standard t-SNE cannot visualize uncertainty, where the low dimension embeddings of points are deterministic and do not correspond to its uncertainty in the original space, Ut-SNE captures the uncertainties of the data's structure and provides a probabilistic visualization of each point as well as the cluster of the point cloud.
    }
 
    \label{fig:toy}
\end{figure}

\subsection{Visualizing the uncertainty of human breast cell}
To illustrate how Ut-SNE uncovers the potentially uncertain structure of the biological dataset, we applied Ut-SNE to a real-world scRNA-seq dataset of human breast tissue. This dataset comprises approximately 430,000 cells from 69 different surgical tissue samples. The original study investigated the heterogeneity of breast tissues, showing distinct changes in normal, preneoplastic, and tumor states.  This revealed significant alterations in the cellular composition of breast tissues have potential relevance for cancer immunotherapies. We seek to determine whether our methods could better comprehend the transcriptomic system, considering uncertainty, compared to current methods. 
At first, we follow the methods outlined in \cite{kobak2019art} to preprocess the dataset. Then we perform Ut-SNE on the processed dataset with uncertainty to capture the inherent structure. 
To effectively compare the visualizations between Ut-SNE and standard t-SNE, we opt for transcriptomic types exhibiting discernible clusters as generated by t-SNE, which implies a significant level of differentiation among these types. 
For both Ut-SNE and standard t-SNE, we choose two different methods to initialize the embedding position of the samples, such as random initialization and initialization based on PCA. In particular, Ut-SNE focuses on incorporating uncertainty in the computation of pair distance, wishing to consider various probabilities of the samples' position. However, standard t-SNE aims at computing the deterministic pairwise distance in a scalar space, which is sensitive to noise caused by data collection and experiment variability. 

To evaluate the quality and reliability of the embeddings generated by standard t-SNE and Ut-SNE, we use three metrics from \cite{kobak2019art}: $\mathrm{KNN}$, $\mathrm{KNC}$, and $\mathrm{CPD}$. Specifically, $\mathrm{KNN}$ measures the proportion of $k$-nearest neighbors in the high-dimensional space that are accurately preserved in the low-dimensional embedding, reflecting the retention of local (microscopic) structures. $\mathrm{KNC}$ quantifies the proportion of retained $k$-nearest neighbors for class centroids between the high- and low-dimensional spaces, focusing on the preservation of class-relative positions and medium-scale structures. $\mathrm{CPD}$ evaluates global structure preservation by calculating the Spearman correlation between pairwise distances in the high- and low-dimensional spaces. A high $\mathrm{CPD}$ score indicates strong preservation of global structure.

In \fref{breast}, we use these metrics to measure the quality of the embedding of the breast cell dataset. Both Ut-SNE and standard t-SNE using PCA initialization perform much better than using random initialization in preserving the local and global structure. Comparison of Ut-SNE's and standard t-SNE's visualizations revealed several key differences in the quality of the embeddings. When using the initialization based on PCA, Ut-SNE consistently generates higher-quality embeddings than standard t-SNE in preserving local, medium-scale, and global structures.
Additionally, the visualization using Ut-SNE with random initialization demonstrates better preservation of global structures in the data, leading to a more faithful representation of the underlying relationships among data points. Standard t-SNE resulted in more scattered and confused groupings, especially for clusters with subcluster structures. As shown in subplot (a), a few samples in the gray-blue cluster (indicating $T$ cells) were significantly distinct from the rest of the gray-blue subcluster, distorting the cell hierarchy. This conclusion of the visualization is quite similar to 
the analysis of original research.

\begin{figure}[h!]
    \centering
    \renewcommand{\tabcolsep}{1.0mm}
    \scriptsize
    \begin{tabular}{p{0.5mm}*{2}{c}}
        & \figclip{10 0 10 2}{0.45}{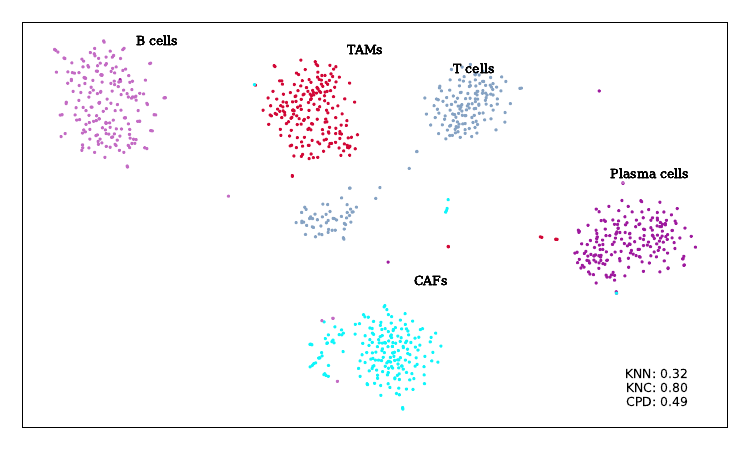}
        & \figclip{10 0 10 2}{0.45}{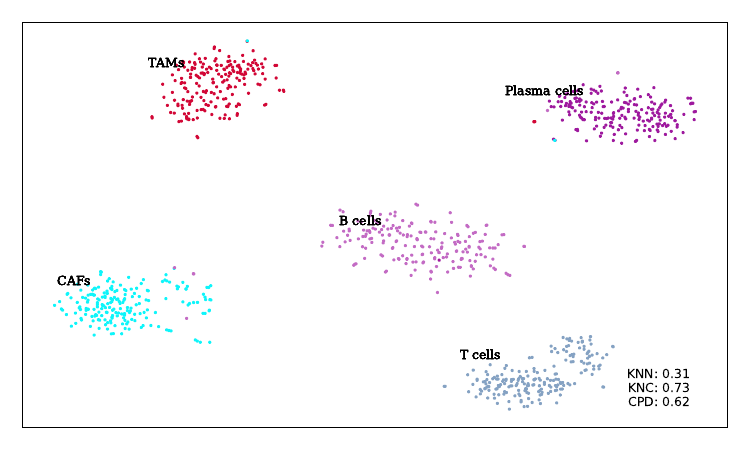} \\
        & (a) & (b) \\
        & \figclip{10 0 10 2}{0.45}{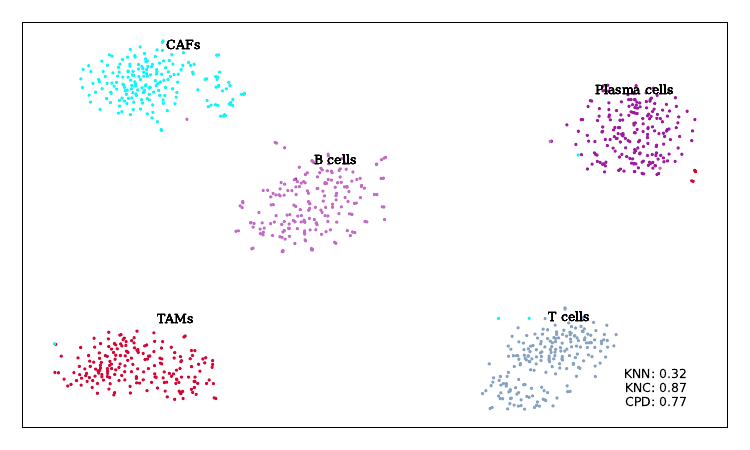}
        & \figclip{10 0 10 2}{0.45}{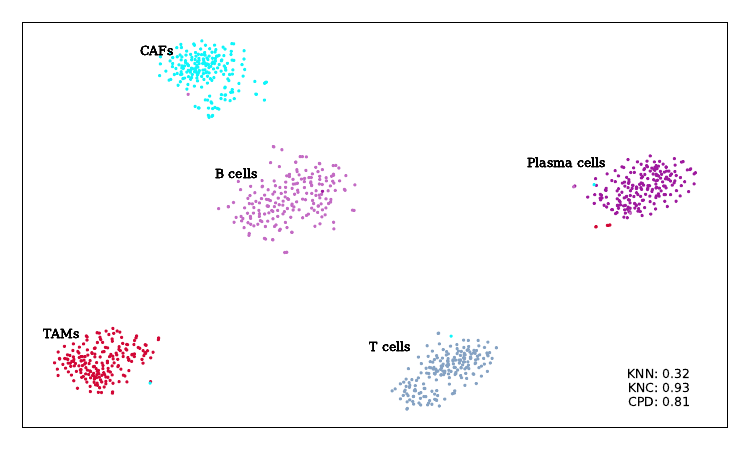}   \\
        & (c) & (d)
    \end{tabular}
    \caption{Visualization on the embedding of breast cell dataset. We label clusters with cell types including $B$ cell (in magenta), TAM (in red), $T$ cell (in gray-blue), CAF (in cyan), and plasma cell (in purple).  These visualizations included: (a) standard t-SNE with the random initialization; (b) Ut-SNE with random initialization; (c) standard t-SNE with PCA initialization; and (d) Ut-SNE with the same PCA initialization. While standard t-SNE often produces scattered visualizations, where the apparent size of a point cluster (distinguished by different colors) does not reflect the space it occupies in the original data, Ut-SNE more accurately captures the true structure by incorporating uncertainty information.
    }
    \label{fig:breast}
\end{figure}

\subsection{Exploring the single-cell of human adipose cell}
Adipocytes have multiple functions that affect the body's overall metabolism. They store excess energy as triglycerides and release it as needed, thus playing a crucial role in the metabolic function of adipose tissue. However, the study of adipogenesis often uses bulk approaches that tend to overlook the differences between individual cells, potentially leading to inaccurate conclusions.  scRNA-seq has gained significant traction in adipogenesis research. While major cell types are generally discerned, the uncertainty at the single-cell level is caused by humans or technology across the whole sequencing process.  

In this experiment, we aim to study the heterogeneity of cells in their differentiation process from multiple aspects. The scRNA-seq dataset was gathered both before and during the adipogenic differentiation of SGBS cells to allow for analyses of the heterogeneity of the transcriptional states before and during adipogenesis. It is important to consider the technical variation and other impact factors. Every cell in this experiment begins life as fat-specific progenitor cells(preadipocytes) showing increased claudin 11 (CLDN11); these cells covert into the differentiating cells or adipocytes. 

In \fref{adipocyte}, we present the embeddings of single-cell samples with uncertainty, utilizing various well-established dimensionality reduction techniques. While all methods generally cluster the data corresponding to different cell types, notable differences emerge among them. Specifically, the data points are dispersed across the two-dimensional plane, particularly evident in the preadipocyte cell samples (represented by the blue clusters) within the embeddings generated by MDS, Isomap, and standard t-SNE. Furthermore, the cell data structure is inaccurately projected into a linear or spherical form rather than reflecting the typical strip shape observed in the original study \cite{li2023single},  failing to adequately capture its inherent complexity and multidimensional structural characteristics, as illustrated in subplots (a, b, d).
In contrast, Ut-SNE effectively produces a pronounced clustering effect, yielding a more compact data distribution while significantly reducing the distortion of high-dimensional data in the low-dimensional space.

Subplot (b) shows the closer strip view of the data points, which highlights the relationship between the variables more clearly. 
Then, we focus here on the single cluster, which stands out as remarkably distinct between the two visualizations. Although both gather the same type of cells, Ut-SNE predicts the cluster's distribution of them, with a single cell's distribution projected in the low-dimensional space \fref{adipocyte-distribution}. The method is to generate several samples from every cell's distribution in a Monte Carlo fashion, estimating the cluster empirical distribution in every dimension. The strategy is to generate multiple samples from each cell's distribution in a Monte Carlo manner, thereby estimating the empirical distribution of the clusters in each dimension.  
Compared to subplots (b) and (d) of \fref{adipocyte-distribution}, the preadipocyte cluster is far away from the adipocyte cluster, located at the low density of the adipocyte distribution. This indicates that there is a clear heterogeneity between preadipocytes and mature adipocytes. 

\begin{figure}[h!]
    \centering
    \renewcommand{\tabcolsep}{1.0mm}
    \scriptsize
    \begin{tabular}{p{0.35mm}*{2}{c}}

        & \figclip{0 30 0 30}{0.40}{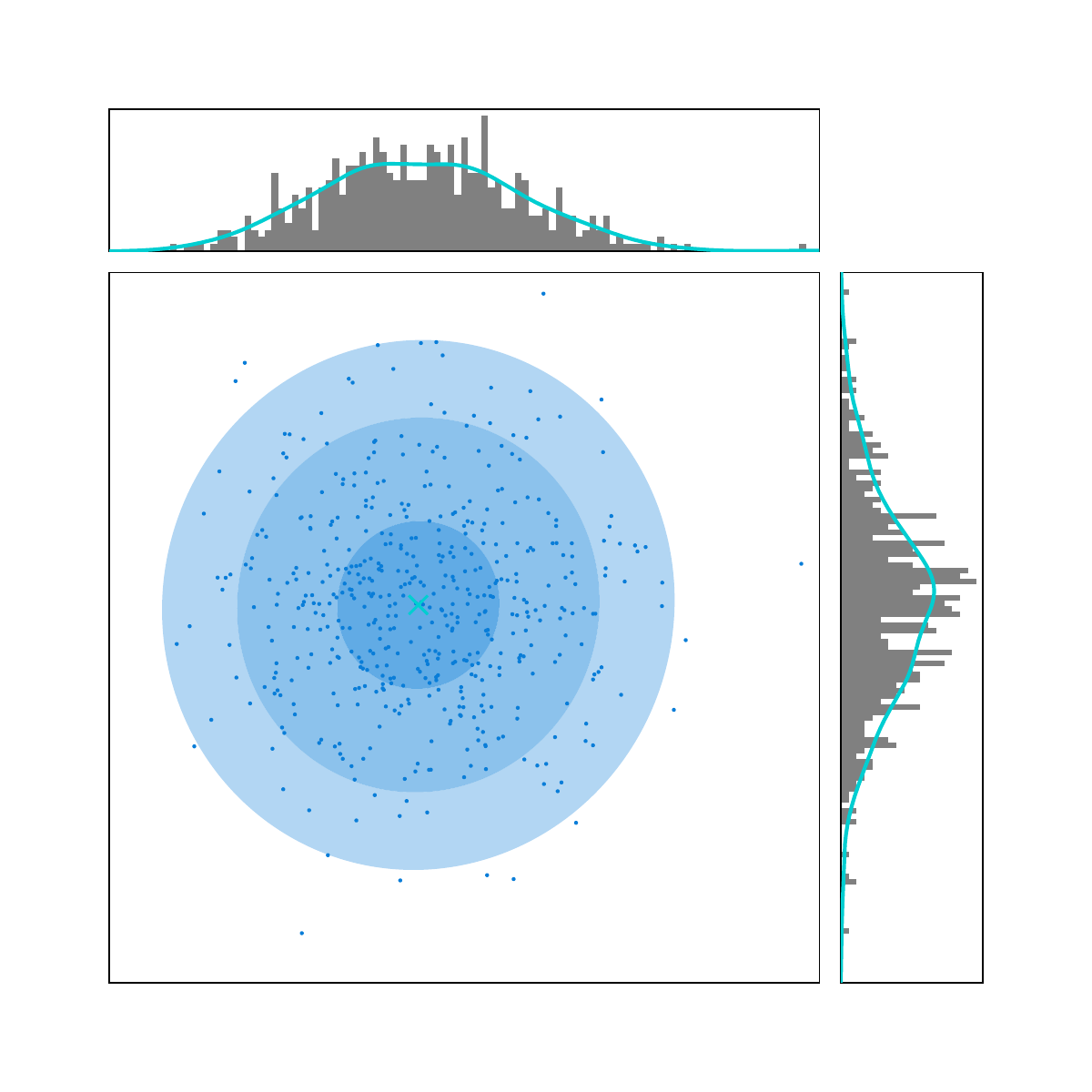}
        & \figclip{0 0 0 2}{0.35}{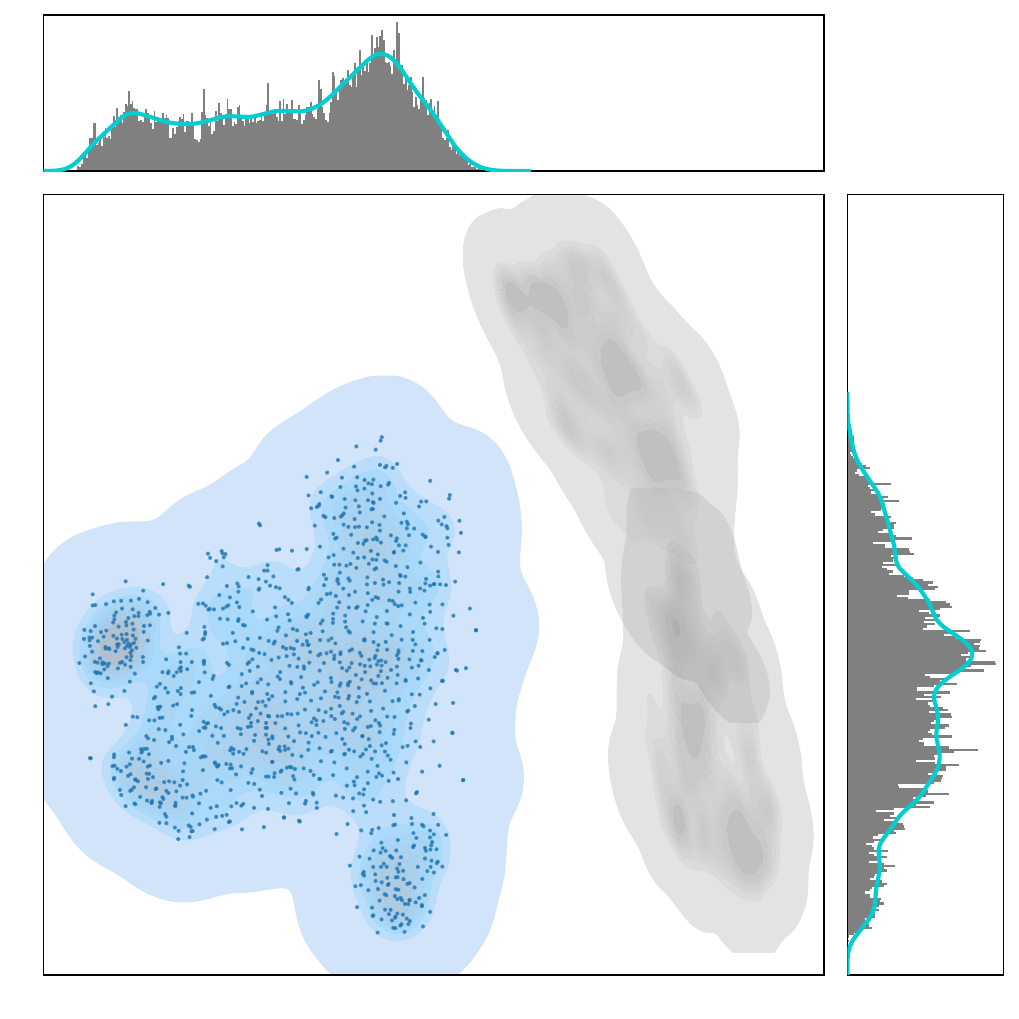}   \\
        & (a) & (b) \\
        & \figclip{0 0 0 2}{0.34}{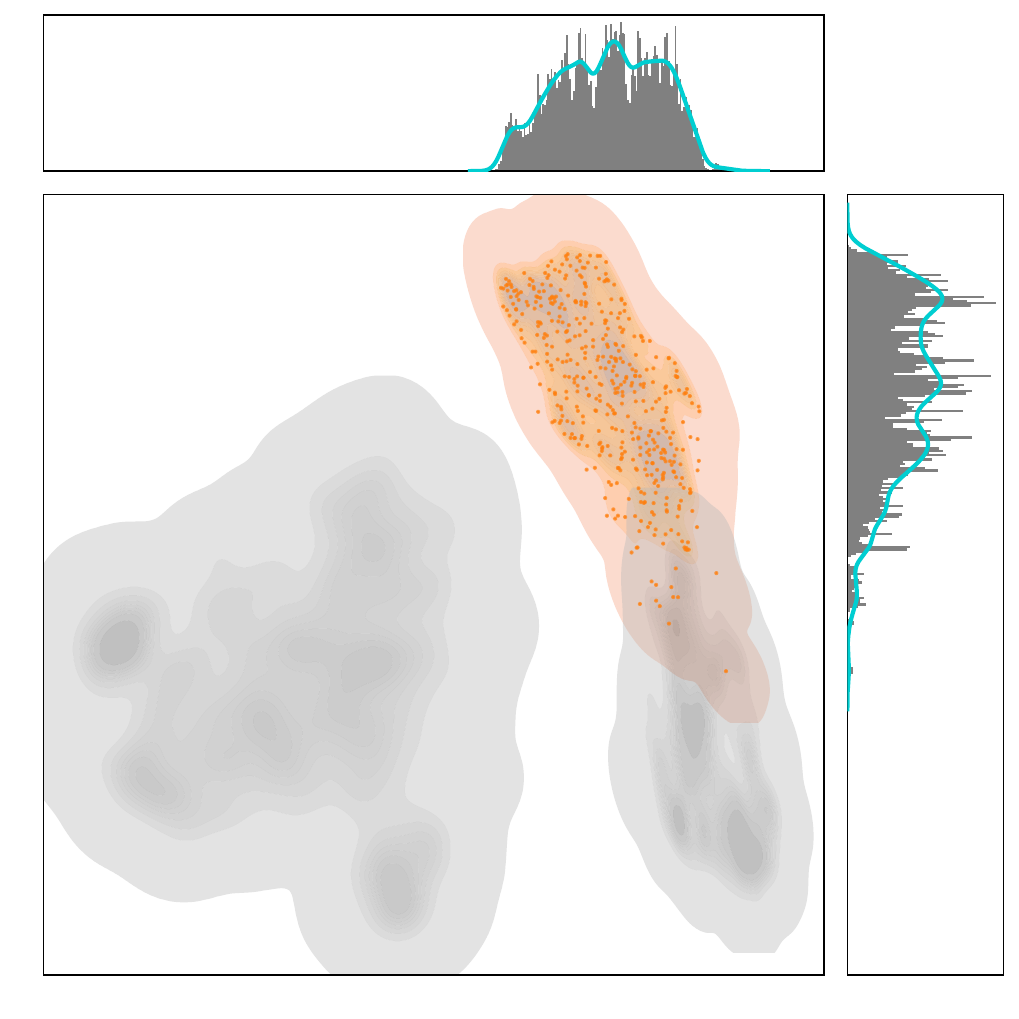}
        & \figclip{0 0 0 2}{0.34}{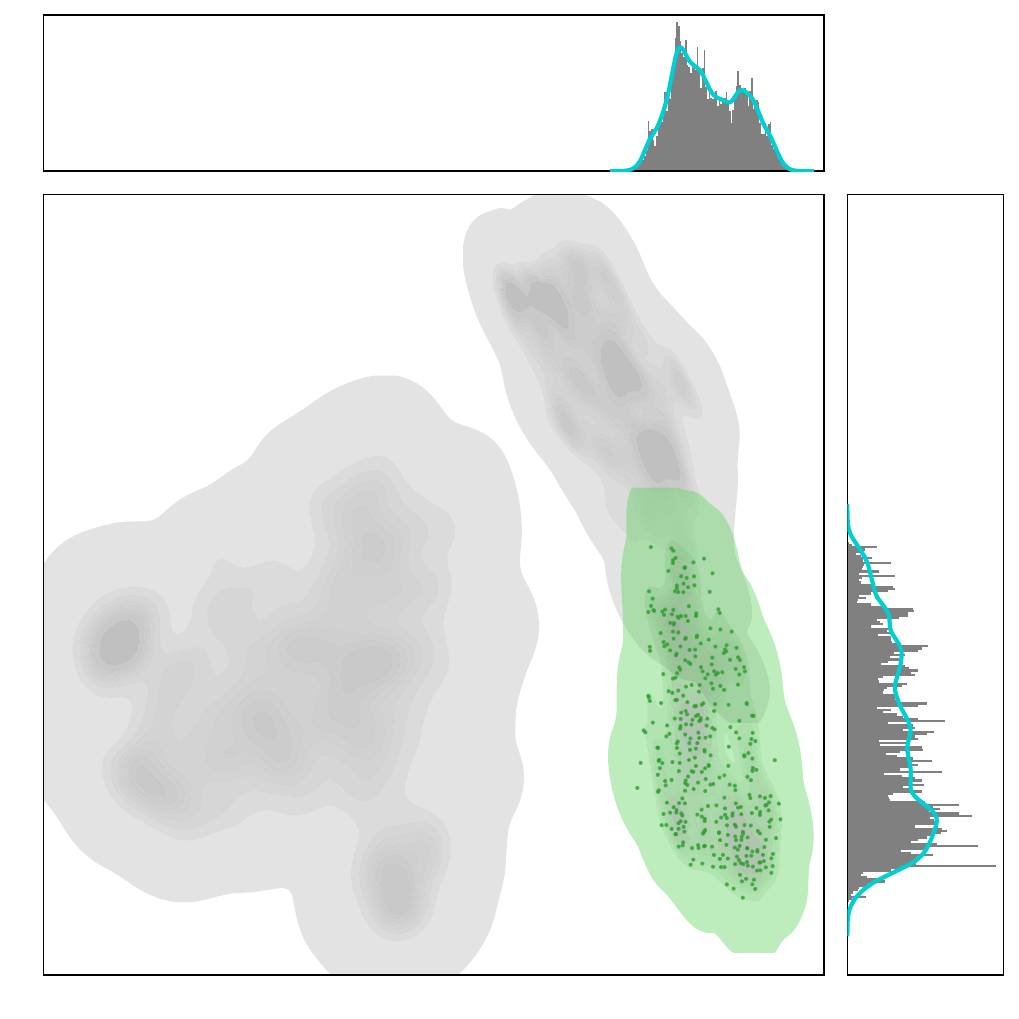}   \\
        & (c) & (d) \\        
        & \figclip{0 0 0 2}{0.35}{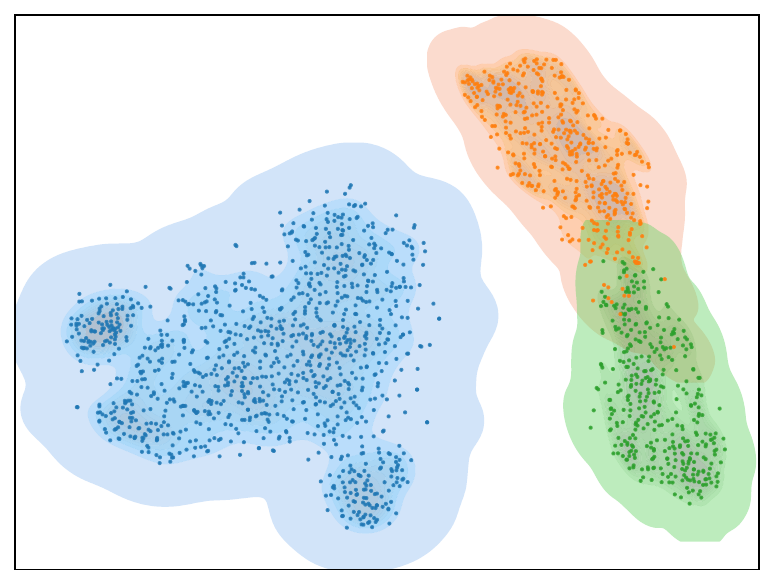}\\
        & (e) \\
    \end{tabular}
    \caption{Visualization on the distribution of adipocyte cell dataset. In contrast to standard t-SNE, Ut-SNE provides a probabilistic representation of the low-dimensional embedding. In subplot (b) (and subsequently) the uncertainty of the low-dimensional embedding is visualized by varying the intensity of the background pixels. The darker the pixel the higher the precision of the mapping.
    Subplot (b) visualizes the embeddings of the preadipocytes. 
    Every preadipocyte cell has an underlying distribution that can be projected as subplot (a) (For simplicity, we show the sampled embeddings). 
    Subplots (c) and (d) visualize the distributions of different cells, respectively.
    }\label{fig:adipocyte-distribution}    
\end{figure}

\begin{figure}[h!]
    \centering
    \renewcommand{\tabcolsep}{1.0mm}
    \scriptsize
    \begin{tabular}{p{0.5mm}*{2}{c}}
        & \figclip{2 0 6 2}{0.83}{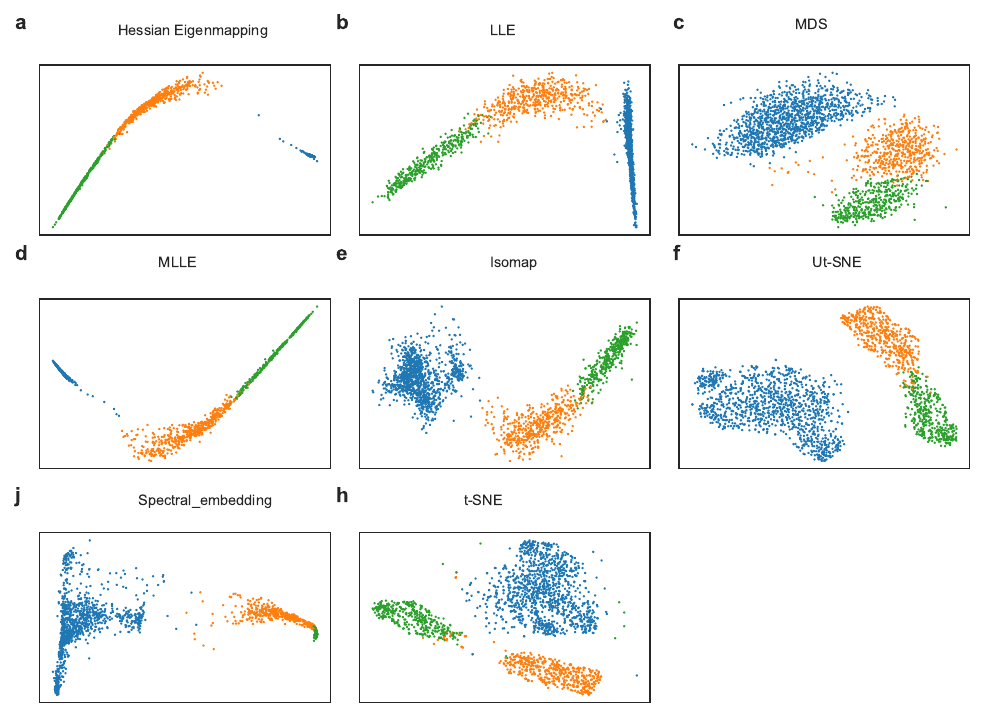}
    
    \end{tabular}
    \caption{Visualization on the embedding of adipocyte cell dataset. 
    }
    \label{fig:adipocyte}
\end{figure}

\section{Discussion and Conclusion}\label{sec12}
Ut-SNE is an efficient statistical method for visualizing single-cell transcriptomics with uncertain information. our method overcomes a major limitation of traditional dimensionality reduction techniques, which neglects the uncertainty of gene expression across single-cell transcriptomics.  Ut-SNE merges uncertain information based on the probability space to compute distances between samples. Based on the distance matrix containing uncertain information, Ut-SNE takes into account the similarities and differences in the data points with uncertainty, ultimately creating a more accurate representation of the relationships between them. Although t-SNE reveals clustering patterns and potential structures, we have demonstrated across a range of datasets that the uncertain information incorporated into our visualizations reveals insights that can enhance understanding of biology beyond what existing visualization tools offer.  Our Ut-SNE are broadly applicable to other visualization algorithms, including recent extensions of t-SNE and other embedding.

Our Ut-SNE underwent extension experiments across diverse visualization and analysis tasks of high-dimensional sequential scRNA-seq data, achieving information incorporation capacity that surpasses or rivals existing t-SNE designed for single data messages. These experiments underscore the broad applicability and milestone of scRNA-seq data. Ut-SNE shows significant promise for biological data analysis, with previous applications encompassing human breast tissue, adipose tissue, and immune cells. The fact that standard t-SNE struggles to visualize the uncertain information contained is one of the well-known limitations.  The embedding is structured to show the corresponding uncertain information while preserving the data structure. Ut-SNE smartly takes into account uncertain information by integrating knowledge of uncertainty with data, allowing for the effortless identification of shifts in variability and the formulation of biological hypotheses. 

However, despite its promising capabilities, Ut-SNE is not without potential limitations and areas for improvement.

\bibliographystyle{unsrt}  
\bibliography{Ut-SNE}

\begin{thebibliography}{10}

\bibitem{svensson2018exponential}
Valentine Svensson, Roser Vento-Tormo, and Sarah~A Teichmann.
\newblock Exponential scaling of single-cell {RNA}-seq in the past decade.
\newblock {\em Nature protocols}, 13(4):599--604, 2018.

\bibitem{ranek2023feature}
Jolene~S Ranek, Wayne Stallaert, Justin Milner, Natalie Stanley, and Jeremy~E Purvis.
\newblock Feature selection for preserving biological trajectories in single-cell data.
\newblock {\em bioRxiv}, 2023.

\bibitem{mackiewicz1993principal}
Andrzej Ma{\'c}kiewicz and Waldemar Ratajczak.
\newblock Principal components analysis ({PCA}).
\newblock {\em Computers \& Geosciences}, 19(3):303--342, 1993.

\bibitem{kurita2019principal}
Takio Kurita.
\newblock Principal component analysis ({PCA}).
\newblock {\em Computer Vision: A Reference Guide}, pages 1--4, 2019.

\bibitem{cai2022theoretical}
T~Tony Cai and Rong Ma.
\newblock Theoretical foundations of t-{SNE} for visualizing high-dimensional clustered data.
\newblock {\em Journal of Machine Learning Research}, 23(301):1--54, 2022.

\bibitem{van2008visualizing}
Laurens Van~der Maaten and Geoffrey Hinton.
\newblock Visualizing data using t-{SNE}.
\newblock {\em Journal of machine learning research}, 9(11), 2008.

\bibitem{roweis2000nonlinear}
Sam~T Roweis and Lawrence~K Saul.
\newblock Nonlinear dimensionality reduction by locally linear embedding.
\newblock {\em science}, 290(5500):2323--2326, 2000.

\bibitem{li2008locally}
Bo~Li, Chun-Hou Zheng, and De-Shuang Huang.
\newblock Locally linear discriminant embedding: {A}n efficient method for face recognition.
\newblock {\em Pattern Recognition}, 41(12):3813--3821, 2008.

\bibitem{carroll1998multidimensional}
J~Douglas Carroll and Phipps Arabie.
\newblock Multidimensional scaling.
\newblock {\em Measurement, judgment and decision making}, pages 179--250, 1998.

\bibitem{saeed2018survey}
Nasir Saeed, Haewoon Nam, Mian Imtiaz~Ul Haq, and Dost~Bhatti Muhammad~Saqib.
\newblock A survey on multidimensional scaling.
\newblock {\em ACM Computing Surveys (CSUR)}, 51(3):1--25, 2018.

\bibitem{hagele2022uncertainty}
David H{\"a}gele, Tim Krake, and Daniel Weiskopf.
\newblock Uncertainty-aware multidimensional scaling.
\newblock {\em IEEE Transactions on Visualization and Computer Graphics}, 29(1):23--32, 2022.

\bibitem{wickramasinghe2019deep}
Chathurika~S Wickramasinghe, Kasun Amarasinghe, and Milos Manic.
\newblock Deep self-organizing maps for unsupervised image classification.
\newblock {\em IEEE Transactions on Industrial Informatics}, 15(11):5837--5845, 2019.

\bibitem{mcinnes2018umap}
Leland McInnes, John Healy, and James Melville.
\newblock Umap: Uniform manifold approximation and projection for dimension reduction.
\newblock {\em arXiv preprint arXiv:1802.03426}, 2018.

\bibitem{becht2019dimensionality}
Etienne Becht, Leland McInnes, John Healy, Charles-Antoine Dutertre, Immanuel~WH Kwok, Lai~Guan Ng, Florent Ginhoux, and Evan~W Newell.
\newblock Dimensionality reduction for visualizing single-cell data using {UMAP}.
\newblock {\em Nature biotechnology}, 37(1):38--44, 2019.

\bibitem{belkin2003laplacian}
Mikhail Belkin and Partha Niyogi.
\newblock Laplacian eigenmaps for dimensionality reduction and data representation.
\newblock {\em Neural computation}, 15(6):1373--1396, 2003.

\bibitem{levin2016laplacian}
Keith Levin and Vince Lyzinski.
\newblock Laplacian eigenmaps from sparse, noisy similarity measurements.
\newblock {\em IEEE Transactions on Signal Processing}, 65(8):1988--2003, 2016.

\bibitem{kobak2019art}
Dmitry Kobak and Philipp Berens.
\newblock The art of using t-{SNE} for single-cell transcriptomics.
\newblock {\em Nature Communications}, 10(1):5416, 2019.

\bibitem{van2021compression}
Scott Van~Buren, Hirak Sarkar, Avi Srivastava, Naim~U Rashid, Rob Patro, and Michael~I Love.
\newblock Compression of quantification uncertainty for sc{RNA}-seq counts.
\newblock {\em Bioinformatics}, 37(12):1699--1707, 2021.

\bibitem{li2023single}
Jiehan Li, Christopher Jin, Stefan Gustafsson, Abhiram Rao, Martin Wabitsch, Chong~Y Park, Thomas Quertermous, Joshua~W Knowles, and Ewa Bielczyk-Maczynska.
\newblock Single-cell transcriptome dataset of human and mouse in vitro adipogenesis models.
\newblock {\em Scientific Data}, 10(1):387, 2023.

\end{thebibliography}

\end{document}